\documentclass[twocolumn,aps,showpacs,groupedaddress,floatfix,prl]{revtex4-1}
\usepackage{amsmath}
\usepackage{graphicx}
\usepackage{dcolumn}
\usepackage{bm}
\usepackage{verbatim}
\usepackage{color}
\usepackage{hyperref}
\usepackage{ulem}
\begin{document}
\setlength{\parskip}{0mm}
\title{Entropy Scaling Laws in Self Propelled Glass Formers}
\author{Sachin C.N. and Ashwin Joy}
\email{ashwin@iitm.ac.in}
\affiliation{Department of Physics, Indian Institute of Technology Madras, Chennai - 600036, India}
\date{\today}
\begin{abstract}
  Predicting transport from equilibrium structure is a challenging problem in liquid state physics.
  Here we probe a glass forming liquid composed of self-propelled ``active'' particles and show that
  increasing the duration of self-propulsion $\tau_p$ makes the pair excess entropy $S_2$ more
  negative, thereby reducing the number of accessible configurations per particle.  At moderate
  values of effective temperature $T$, the self-diffusivity is Arrhenius and in a reduced form obeys
  a Dzugutov like scaling law $D^* \sim e^{\alpha S_2}$, directly yielding us the scaling formula
  $S_2 \sim -1/T$. In the strongly super-cooled regime, Dzugutov law does not apply and the entropy
  follows a power law $S_2 \sim -1/T^{\beta}$ all the way up to the glass transition $T_g$. To
  demonstrate generality, we set the particle interactions to be purely repulsive (PR) in one case
  and Lennard-Jones (LJ) in the other, and find that in both the cases, the reported scaling laws
  are robust over three decades of variation in $\tau_p$.  Our results may apply to transport in
  active colloidal suspensions, passive tracers in bacterial baths, and self-propelled granular
  media, to mention a few.
\end{abstract} 

\maketitle
Liquids in which the constituent particles are self-propelled or ``active'' can display a class of collective behavior that is typically not observed
in conventional systems\cite{LBerthier, gonzalez2012soft, Theurkaff, PhysRevLett.109.248109}. The coherent motion of these active particles has been
shown to act as a precursor to flocking- an exotic ordered phase that arises when the mean velocity $\langle \vec{v} \rangle \neq 0$, with examples
ranging from bird flocks and insect swarms to granular matter and dense colloids \cite{PhysRevE.58.4828, toner2005hydrodynamics,
marchetti2013hydrodynamics, cavagna2014bird, Kaisere1601469}. Other notable mentions where self-propulsion can profoundly affect liquid dynamics are
jamming \cite{bechinger2016active}, phase separation \cite{PhysRevLett.110.055701,fily2014freezing} and phase transitions \cite{czirok1999collective,
fily2012athermal}. It is therefore natural to ask how activity modulates the dynamics and structure of a liquid especially at low temperatures where
collective behavior is dominant. An important problem that has recently witnessed a sharp surge of interest is to establish the role of
self-propulsion in the dynamics of super-cooled liquids nearing the glass transition\cite{JuanRuben}. Preliminary works in this area remain
inconclusive on whether self-propulsion mitigates \cite{mandal2016active} or enhances \cite{flenner2016nonequilibrium} sluggish dynamics near a glass
transition. The role of ``activity" in modifying the potential energy landscape may help in understanding these contradicting claims
\cite{ni2013pushing}. The presence of an equilibrium counterpart or lack thereof in the sluggish dynamics of these self propelled liquids therefore
presents as an important research direction. Our letter focuses on this direction and reports a detailed investigation of the
structure and transport in a model active glass forming liquid. In what follows, we will provide robust scaling laws for pair excess
entropy - a quantity that is directly amenable in particle resolved experiments\cite{yokoyama2002excess}. 
We use the following numerical model of a ``living'' fluid where the governing equation for the i$^{\text{th}}$ particle, reads as-
\begin{align}
  \bm{\dot{r}}_i &=   \frac{1}{m\gamma} \biggl(- \nabla_i U + \bm{f}_i\biggr) \nonumber \\ 
  \bm{\dot{f}}_i &= \frac{1}{\tau_p} \biggl(- \bm{f}_i + \sqrt{2 m \gamma k_B T} \;\bm{\eta}_i \biggr) 
  \label{AOUP}
\end{align}

Put simply, the particle dynamics is governed by an overdamped Ornstein-Uhlenbeck type of stochastic process that has been used
extensively to model athermal active fluids \cite{PhysRevE.91.042310,koumakis2014directed,PhysRevE.91.062304,marconi2015towards,FelixGinot}.  In this
model, self-propulsion is described completely using only two parameters, namely, the effective temperature $T$ and a certain time scale $\tau_p$- the
former manifesting in the strength of the self-propulsion force $\bm{f}_i$ and the latter indicates the duration of $\bm{f}_i$. We take $\bm{\eta}_i$
to be a Gaussian white noise with zero mean and unit variance. The parameter $\gamma$ refers to the friction coefficient and $U$ denotes the potential
energy due to particle interactions. In this letter, we have used this model for self propulsion in a well known glass former and will report scaling
laws for pair excess entropy and self diffusivity in the super-cooled regime. It should be noted that the Eq. \ref{AOUP} reduces to the standard
overdamped Brownian dynamics as $\tau_p \rightarrow 0$ and has been shown to satisfy detailed balance in the small $\tau_p$ limit
\cite{PhysRevLett.117.038103}. The friction coefficient $\gamma$ is set to unity in all the runs and Eq. \ref{AOUP} is integrated using a fully
implicit scheme and a time step of $10^{-4}$ \cite{PhysRevA.40.3381}. In the following we provide the details of our numerical work.

We have used the Kob-Andersen binary glass forming liquid with $80\%$ large $(L)$ and $20\%$ small $(S)$
particles interacting via the Lennard-Jones (LJ) potential energy\cite{KOBANDERSONLJPOTENTIAL}
\begin{equation}
  U_{\text{LJ}} = \sum _{i<j} 4 \epsilon_{ij} \biggl [\biggl(\frac{\sigma_{ij}}{r_{ij}}\biggr)^{12} - \biggl(\frac{\sigma_{ij}}{r_{ij}}\biggl)^6 \biggr].
  \label{KA}
\end{equation}
To demonstrate generality, we repeated simulations on a purely repulsive (PR) potential energy\cite{pedersen2010repulsive}
\begin{equation}
  U_{\text{PR}}  = \sum_{i<j} 1.945 \epsilon_{ij} \biggl(\frac{\sigma_{ij}}{r_{ij}}\biggr)^{15.48}
  \label{}
\end{equation}
and observed qualitatively similar results. For both types of interactions, we took $\epsilon_{LL}$,
$\sigma_{LL}$ and $\sqrt{m \sigma_{LL}^2 /
\epsilon_{LL}}$ as the units of energy, length and time respectively. In these units, the potential parameters become $\epsilon_{SS} = 0.50, \epsilon_{LS} = 1.50, \sigma_{SS} =
0.88$, and $\sigma_{LS} = 0.80$.
\begin{figure}[h]
        \centering
        \includegraphics[width=\textwidth,height=0.25\textheight,keepaspectratio]{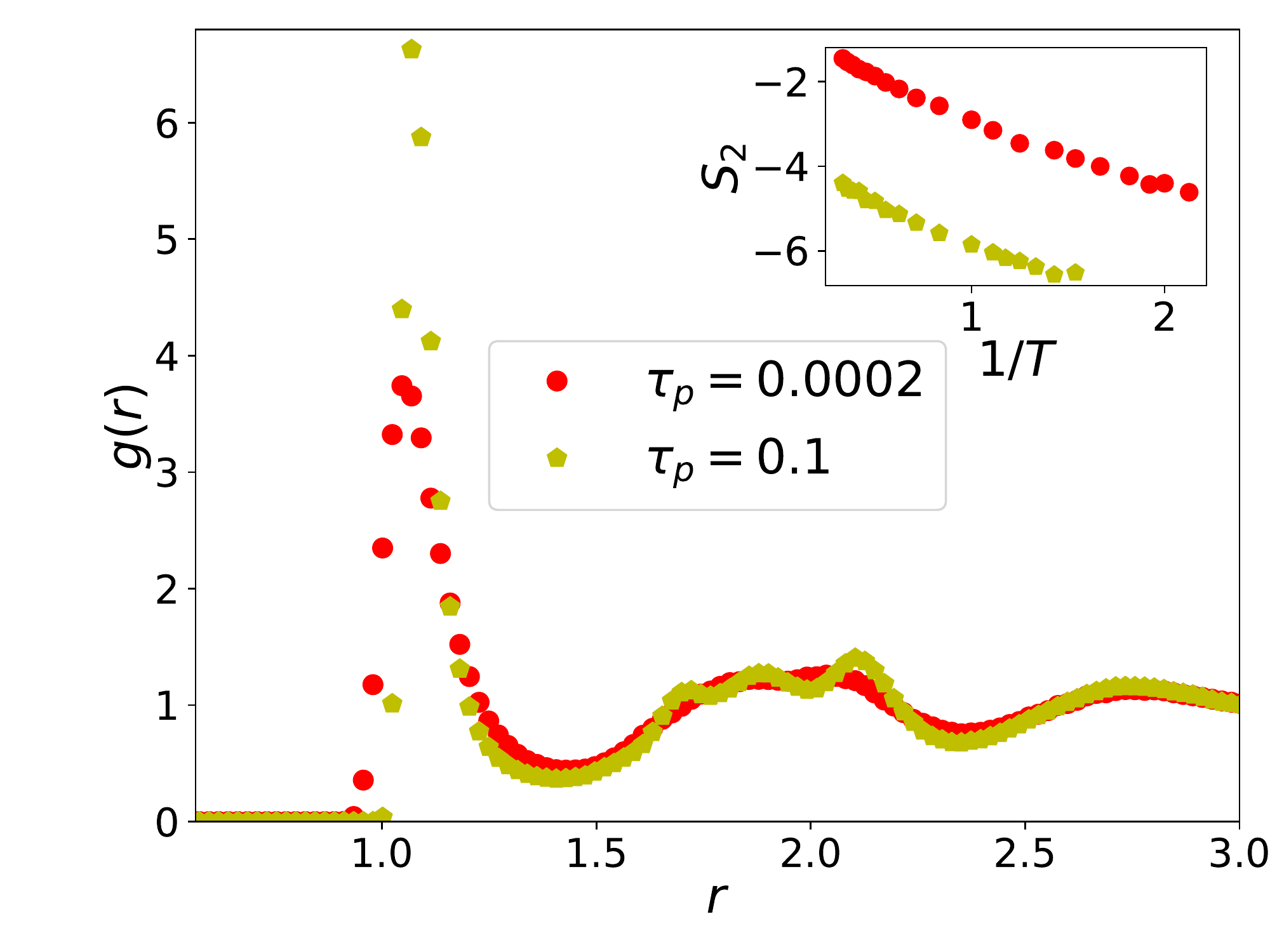}
        \caption{(color online). Decrease in pair excess entropy $S_2$ is
                  correlated with increase in short range order. Inset: For a single temperature $T = 0.70$, we show that the short range
                  order is correlated with $\tau_p$, and this is true for all $T$.}
        \label{fig-s2gr}
\end{figure}
In order to derive our scaling laws, we need to first set up the connection between the dynamics and the underlying structure in our model active
liquid. To that end, we have made extensive measurements on pair excess entropy $S_2$ (explained below) that essentially captures the correction to
the ideal gas entropy due to pair correlations. It is then straightforward to compute $S_2$ using the prescription by Wallace \cite{Wallace87} -
\begin{equation}
  S_{2} = -2 \pi \sum_{\mu} \chi_\mu \sum_{\nu} \rho_\nu \int [1 + g_{\mu\nu}(r) \{\text{ln}\; g_{\mu\nu}(r) -1 \}] r^2\; \text{d} r  \label{S_2}
\end{equation}

\begin{figure}[h]
  \centering
  \includegraphics[width=\textwidth,height=0.25\textheight,keepaspectratio]{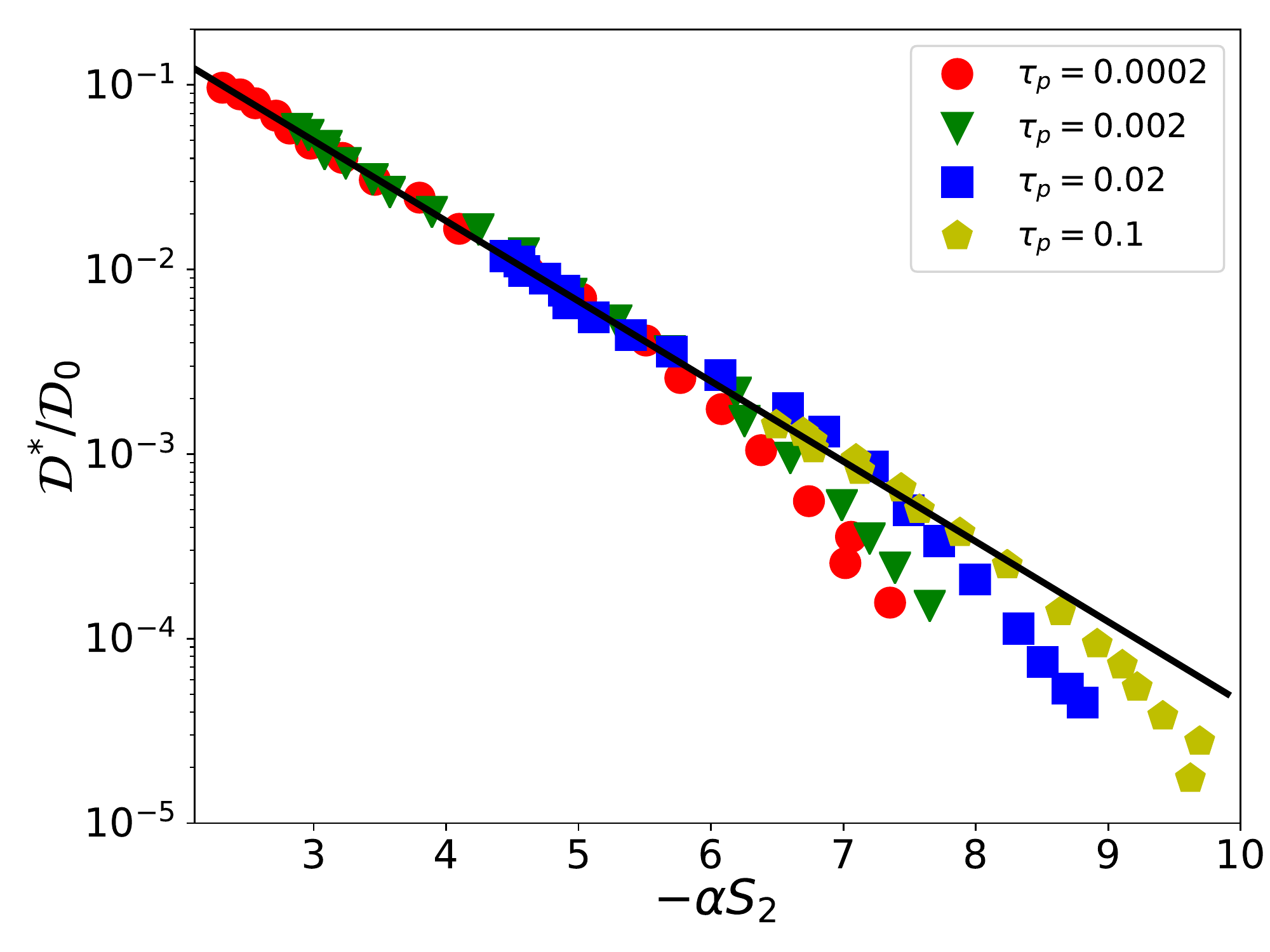}
  \caption{A Dzugutov like scaling law $D^* = D_0 \;\text{exp}\;(\alpha S_2)$ \cite{dzugutov1996universal} applies well at moderate super-cooling for all $\tau_p$. The prefactors 
vary as $D_0 \in [0.21069, 0.974678]$ and $\alpha \in [1.59308,1.47573]$, with the latter showing only a weak dependence on $\tau_p$. Clearly the
scaling law breaks down in strongly super-cooled regime.}
  \label{fig-DS2}
\end{figure}
where the labels $\{\mu,\nu\}$ can refer to the particle types $L$ and $S$. The parameters $\rho_\mu$, $\chi_{\mu}$ and $g_{\mu\nu}(r)$ are
respectively, the partial density, the partial molar fraction and the partial radial distribution function.  Since $S_2$ usually takes up more than
$90 \%$ of the total excess entropy in the liquid state \cite{baranyai1989direct,Ashwin-s2-POP}, we have used $S_2$ throughout our work. As can be
expected, at lower temperatures, this correction $S_2$ becomes more negative due to increasing short range order. In Fig. \ref{fig-s2gr}, we show our
data on $g(r)$ at some effective temperature $T=0.7$ but two different values of persistence time $\tau_p$. At higher $\tau_p$, the first peak of
$g(r)$ becomes taller clearly indicating the growth of short range order \cite{flenner2016nonequilibrium}. As the long range order is only marginally
affected by varying  $\tau_p$, the overall effect of increasing $\tau_p$ is an increase in the magnitude of the area integral mentioned in Eq
\ref{S_2}. This is true at all temperatures and is confirmed in the Fig. \ref{fig-s2gr}:inset. Hence we can conclude that increasing persistence time
has an effect of making the two body excess entropy more negative.

We are now in position to connect structure with liquid diffusivity, and for
this we use the prescription of Hoyt \textit{et al} \cite{hoyt2000test} to obtain a normalized total diffusivity in terms of the scaled contributions
coming from $S$ and $L$ type particles-
\begin{equation}
  \mathcal{D}^* = \biggl(\frac{\mathcal{D}_L}{\Gamma_L}\biggr)^{\chi_L} \biggl(\frac{\mathcal{D}_S}{\Gamma_S}\biggr)^{\chi_S}
  \label{D*}
\end{equation}
where $\mathcal{D}_\mu$ is the partial diffusivity of $\mu$ type particles,
calculated from  mean squared displacement. The scale factors are given by
\begin{equation}
  \Gamma_\mu = 4 \sqrt{\frac{\pi k_B T}{m}} \sum_\nu \sigma_{\mu \nu}^4 g_{\mu \nu}(\sigma_{\mu \nu}) \rho_\nu 
  \label{}
\end{equation}
with $\mu, \nu = \{S, L\}$, as before and $\rho_L = 0.96$, $\rho_S = 0.24$. In Figure \ref{fig-DS2}, we show a plot of $\mathcal{D}^*$
vs. $S_2$ at various values of self propulsion. 
\begin{figure}[h]
  \centering
  \includegraphics[width=\textwidth,height=0.25\textheight,keepaspectratio]{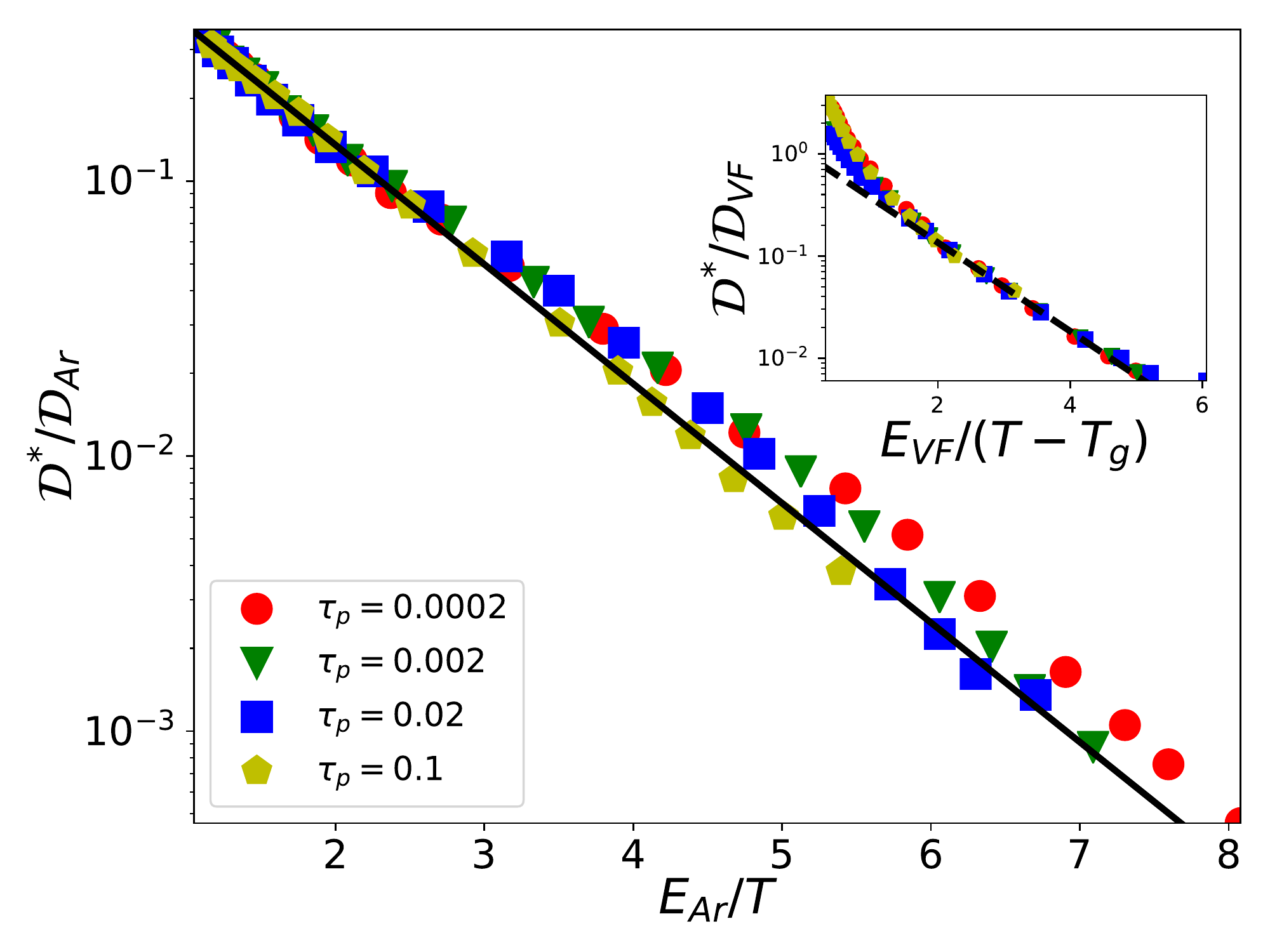}
  \caption{Diffusivity vs. temperature. Data collapse at high temperature is achieved by using the Arrhenius scaling of diffusivity $D$.
    Inset: At lower temperature, a Vogel-Fulcher scaling works well for $D$.}
  \label{fig-DT}
\end{figure}
It is evident from the figure that Dzugutov like scaling law \cite{dzugutov1996universal} 
\begin{equation}
  D^* = D_0 \;\text{exp}\;(\alpha S_2),
  \label{eq-DS2}
\end{equation}
applies in the moderate range of super-cooling at all values of $\tau_p$, the parameters $D_0$ and $\alpha$ being dependent on $\tau_p$. It should be
noted that the scaling law (Eq.\ref{eq-DS2}) does not work at deep super-cooling, possibly due to the breakdown of ergodicity at these temperatures
\cite{dzugutov_ergodicbreaking}. It is important to stress here that at moderate super-cooling, the
entropy scaling law works for over three decades of
variation in $\tau_p$ and therefore allows us to build robust scaling laws for $S_2$ in this regime. To connect $S_2$ with $T$, we turn our attention
to the temperature dependence of diffusivity $D^*$. Figure \ref{fig-DT} shows a plot of $D^*$ vs. $T$ at various values of $\tau_p$ used in our work.
Data collapse at high temperature ($T > 1.0$) is evidently achieved by using the Arrhenius form of diffusivity\cite{wang2015computational}.
\begin{equation}
  D^* = D_{Ar} \;e^{-E_{Ar}/T} \quad (\text{High $T$}).
  \label{eq-Da}
\end{equation}
On the other hand, in the low temperature regime (inset: Figure \ref{fig-DT}), we observe an excellent data collapse using the Vogel-Fulcher (VF) form
\begin{equation}
  D^* = D_{VF} \;e^{-E_{VF}/(T-T_{g})} \quad (\text{Low $T$}),
  \label{eq-Dv}
\end{equation}
We now present a way to extract reliable scaling laws for pair excess entropy $S_2$ that is directly
from particle positions in a typical experiment. In the high temperature regime, we eliminate $D^*$ between equations \ref{eq-DS2} and \ref{eq-Da} to
get a scaling 
\begin{equation}
  S_2 = (-1/\alpha)\;\text{ln}\; (D_0/D_{Ar}) - E_{Ar}/\alpha T \quad (\text{High $T$}).
  \label{eq-S2Hi}
\end{equation}
A similar relation was qualitatively suggested in \cite{LiLiuZhu} in the context of equilibrium passive liquids and we find it remarkable this scaling
law holds for a wide range of $\tau_p$, in our non-equilibrium active liquid. The role of persistence time $\tau_p$ is therefore limited to
modulating the intercept, $(1/\alpha)\text{ln}\; (D_0/D_{Ar})$ and the slope,
$E_{Ar}/\alpha$ of the entropy relation. At low temperatures however, we observe that the entropy
follows a power law 
\begin{equation}
  S_2 = S^g_2 \;(T_g/T)^\beta \quad (\text{Low $T$}),
  \label{eq-S2Lo}
\end{equation}
\begin{figure}
  \includegraphics[width=\textwidth,height=0.25\textheight,keepaspectratio]{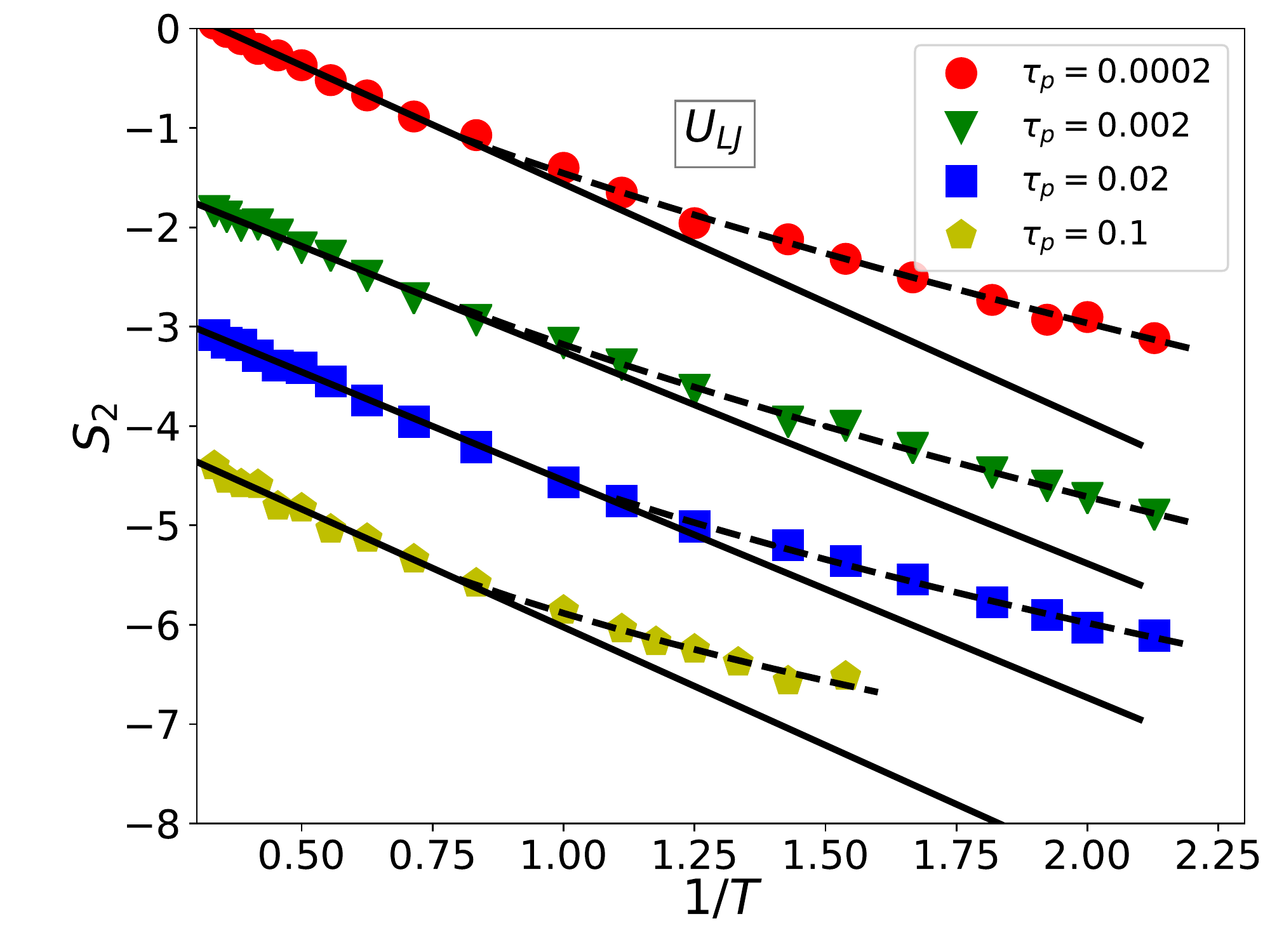}
  \includegraphics[width=\textwidth,height=0.25\textheight,keepaspectratio]{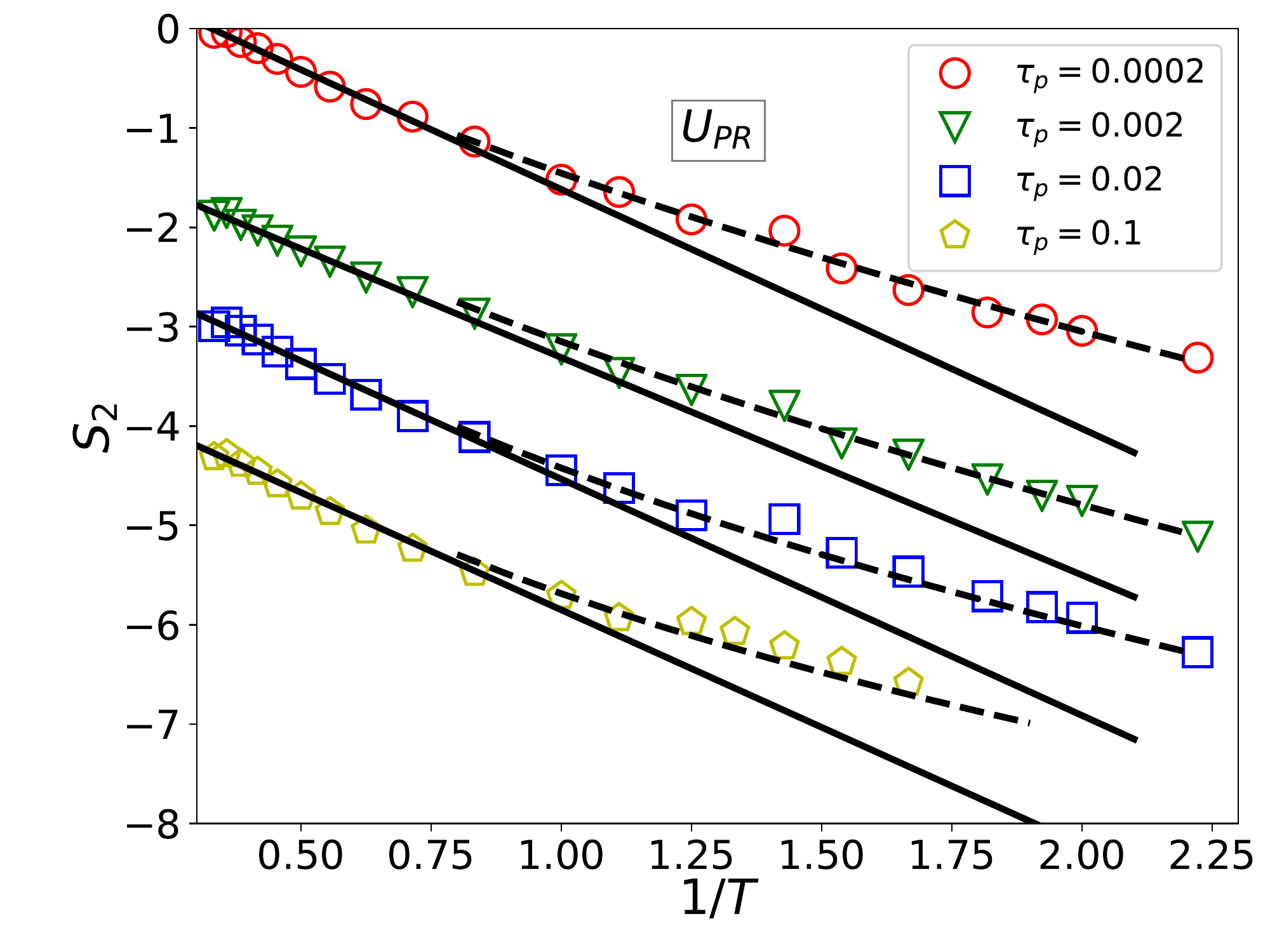}
  \caption{Pair excess entropy $S_2$ vs. effective temperature $T$ at various values of persistence times $\tau_p$. We use filled symbols and empty
    symbols for LJ and PR type particle interactions, respectively.  At high temperatures, the scaling law $S_2 \sim -1/T$ works well with the slope,
    $-E_{Ar}/\alpha$ and the intercept, $(-1/\alpha)\;\text{ln}\; (D_0/D_{Ar})$.  At low
    temperatures however, we find excellent agreement with the power law scaling $S_2 \sim
    -1/T^\beta$. All curves are shifted along $y$-axis for clarity.}
  \label{S2-scaling}
\end{figure}
\par\noindent
all the way up to the lowest temperature where we can achieve an effective equilibrium. The fitting
parameters $T_g$ and $S_2^g$ respectively, the glass transition temperature and the pair excess
entropy at $T_g$. Note $S_2^g <0$ is required to make $S_2 < 0$.  Finally, to demonstrate the
generality of our scaling laws, we plot $S_2$ vs.  $T$ in figure \ref{S2-scaling} for both
Lennard-Jones (LJ) and purely repulsive (PR) particle interactions using filled and empty symbols
respectively. In both the figures, the solid lines demonstrate the high temperature scaling law $S_2
\sim -1/T$ (from Eq.\ref{eq-S2Hi}) with the slope and intercept represented by $-E_{Ar}/\alpha$ and
$(-1/\alpha)\;\text{ln}\; (D_0/D_{Ar})$, respectively.  We can explain this on grounds that the
liquid structure manifests in the number of accessible states available per atom and therefore
strongly affects the rate of cage diffusion. Since cage breaking is necessary for the onset of
diffusive regime, it is natural to expect that reduction in $D$ is concomitant with increasing
$-S_2$. At lower temperatures however, the liquid becomes increasingly non-ergodic and the Arrhenius
behaviour is lost, directly pressing the need for an alternative scaling law. Our data at low
temperatures is in excellent agreement with a power law scaling $S_2 \sim -1/T^\beta$ (from
Eq.\ref{eq-S2Lo}) that remains valid up to the lowest effective temperature accessible to us. A
numerical fit of our data to this power law scaling directly reveals the numerical glass transition
temperature $T_g$. To test the generality of our predictions, we have also verified our scaling laws
with data obtained from a liquid with purely repulsive particle interactions at various values of
persistence time [Ref. Fig. \ref{S2-scaling}]. We therefore assert that these scaling laws are
universal in nature and should be of great utility to soft matter physicists interested in active
glass forming liquids.
  
\textit{Summary:} Our paper reports a careful study of pair excess entropy $S_2$ and its connection
to diffusivity $D$, thus proving the existence of Dzugutov like scaling law in a model active glass
forming liquid. To our knowledge, such studies have been performed only in the domain of equilibrium liquids and no account exists in literature that
deliberates on the nature of this connection in non-equilibrium ``living'' liquids. We focus on a model ``living'' liquid that is essentially far from
equilibrium and where the notion of a effective temperature can be established only in the steady state.  The role of self propulsion in mitigating
transport is carefully examined and is seen to be a precursor for sluggish dynamics. By systematically examining transport and structure, we are able
to predict scaling laws for the pair excess entropy that are both independent of the type of particle interactions, and also valid over a wide range
of persistence time.  As our findings are universal in nature, we believe that they could be of great interest to an experimentalist exploring the transport
phenomena in ``living'' fluids, especially  when  particle positions are resolved at the microscopic level, and $S_2$ becomes directly amenable.

\begin{acknowledgments}
We thank Ethayaraja Mani and Abhijit Sen for discussions and comments on the manuscript. All simulations were done on the VIRGO super cluster of IIT
Madras.
\end{acknowledgments}

\bibliography{manuscript}
\end{document}